\documentclass{aa}  

\usepackage{graphicx}
\usepackage{subcaption}
\usepackage{txfonts}
\usepackage{hyperref}
\usepackage{adjustbox}
\usepackage{booktabs}
\usepackage{tikz}

\begin{document}

\title{Unified Deep Learning Approach for Estimating the Metallicities of RR Lyrae Stars Using light curves from {\it Gaia} Data Release 3}


\author{L. Monti\inst{1}, T. Muraveva\inst{1}, A. Garofalo\inst{1}, G. Clementini\inst{1}, \and M.L. Valentini\inst{2}
}

\institute{INAF - Osservatorio di Astrofisica e Scienza dello Spazio di Bologna, via Piero Gobetti 93/3, 40129 Bologna, Italy;\\ \email{lorenzo.monti@inaf.it} 
}

\date{Received ; accepted }

 
\abstract
{RR Lyrae stars (RRLs) are old population pulsating variables that serve as useful metallicity tracers due to the correlation between their metal abundances and the shape of their light curves. With the advent of ESA’s {\it Gaia} mission Data Release 3 (DR3), which provides light curves for approximately 270,000 RRLs, it has become crucial to develop a machine learning technique that allows for the estimation of metallicities for large samples of RRLs directly from their light curves.}
{We aim to develop and validate a unified Deep Learning (DL) framework capable of accurately estimating metallicities for both fundamental mode (RRab) and first-overtone (RRc) RRLs using $G$-band light curves published in {\it Gaia} DR3. We seek to extend our previous DL model, which was successful for RRab stars, to encompass RRc variables. Our objective is to evaluate the model's performance in terms of accuracy and reliability across both types of RRLs. Through this research, we aim to demonstrate the potential of DL in processing and analyzing large-scale astronomical datasets. Ultimately, we strive to contribute to a more comprehensive understanding of stellar populations and galactic structure through improved metallicity estimations.}
{We employ a Gated Recurrent Units (GRU) based neural network architecture optimized for time-series extrinsic regression. The framework incorporates a rigorous preprocessing pipeline (including phase folding, smoothing, and sample weighting) and is trained using {\it Gaia} DR3 $G$-band light curves and photometric metallicities of RRLs, available in the literature. The model architecture and training implicitly handle the morphological differences between RRab and RRc light curves.}
{Our unified GRU model achieves high predictive accuracy and robust generalization on validation sets for both types of RRLs. For RRab stars, we obtain Mean Absolute Error (MAE) = 0.0565 dex, Root Mean Squared Error (RMSE) = 0.0765 dex, and determination score $R^2$ = 0.9401. For RRc stars, performance is similarly strong with MAE = 0.0505 dex, RMSE = 0.0720 dex, and $R^2$ = 0.9625, demonstrating the framework's effectiveness across different RRL pulsation modes.}
{}



\titlerunning{Unified DL approach for estimating [Fe/H] of RRLs using {\it Gaia} DR3}
\authorrunning{L. Monti, T. Muraveva, A. Garofalo, G. Clementini \and M.L. Valentini}

\maketitle

\section{Introduction}
\label{sec:intro}

RR Lyrae (RRL) stars are low-mass (M < 1 \(M_\odot\)), core helium-burning variables characterized by radial pulsations, with periods typically ranging from 0.2 to 1 day \citep{Smith2004}. Even though recent studies suggest that relatively young and metal-rich RRLs may form through the evolution of close binary systems \citep{Bobrick2024}, the vast majority of RRLs are old (> 10 Gyr), metal-poor stars associated with the Milky Way (MW) halo and old stellar systems such as globular clusters (GCs), dwarf spheroidal (dSph) galaxies and ultra-faint dwarfs (e.g., \citealt{Dallora2006,Clementini2012,Garofalo2013,Sesar2014, Molnar2015, Muraveva2020, Garofalo2021}).
RRLs are classified into three pulsation modes: fundamental mode (RRab), first-overtone (RRc), and double-mode (RRd) stars. Their distinct pulsation characteristics and the occurrence in ancient systems make them powerful tools for studying the structure, formation history, and chemical evolution of the MW and Local Group galaxies \citep{Drake2013, Belokurov2018, Iorio2019, Iorio2021}.



RRL stars serve as useful metallicity ([Fe/H]) tracers. The most direct method for measuring their metal abundances is through high-resolution (HR, R $\geq$ 20,000) spectroscopy, which yields metallicities with an accuracy of $\sim 0.1$ dex but requires significant telescope time. To date, metallicities from HR spectra are available for only a limited number of RRLs (e.g., \citealt{Clementini1995}, \citealt{Nemec2013}, \citealt{Pancino2015}, \citealt{Chadid2017}), although this number has increased 
to a couple of hundreds
in recent years \citep{Crestani2021, Gilligan2021, Dorazi2024}. The metallicities of RRLs can also be measured from low-resolution (LR) spectra using the $\Delta S$ method \citep{Preston1959}, which is based on the ratio of the equivalent widths of Ca and H lines. This approach extends the number of RRLs with available metallicities to thousands (e.g., \citealt{Liu2020}, \citealt{Crestani2021}, \citealt{Fabrizio2021}), though with lower precision (typical uncertainties of $\sim$0.2–0.3 dex).

The metallicities of RRLs can also be determined using only photometric observations, as there is a correlation between RRL metallicity and the shape of the light curve. \citet{Jurcsik1996} found a linear relation between the metallicity of RRab stars and the Fourier parameter $\phi_{31}$ of their light curves in the $V$ band, along with the pulsation period. \citet{Morgan2007} derived a similar relation for RRc stars. Several authors later calibrated these relations in different passbands (e.g., \citealt{Smolec2005}, \citealt{Nemec2013}, \citealt{Iorio2021}, \citealt{Li2023}). These relations offer a pathway to estimate metallicities for large samples of RRLs using only photometric data, thereby bypassing the need for time-intensive spectroscopy. However, accurately calibrating these photometric metallicity relations across different photometric bands has proven challenging. Since HR spectroscopic metallicities are available for only a few hundred RRLs, many calibrations have relied either on LR spectroscopic estimates or on transferring empirical formulae calibrated in one photometric band to another via Fourier parameter transformations \citep{Skowron2016, Clementini2019}. Both approaches introduce potential systematic errors and biases, arising from intermediate calibration steps or from the intrinsic noise and metallicity dependence of the parameter transformations \citep{Dekany2021}. These limitations have often resulted in discrepancies and systematic offsets between metallicity estimates from different methods \citep{Dekany2022}.

A promising solution to this problem is the direct estimation of RRL metallicities from light curves using Machine Learning (ML) and Deep Learning (DL) techniques. 
These approaches allow for modeling non-linear relationships between light-curve morphology and metallicity, potentially bypassing intermediate calibration steps and enabling the direct use of raw time-series data. Predicting metallicity directly from photometric light curves using DL techniques offers several key advantages. First, it enables the analysis of large datasets without the need for time-consuming and costly spectroscopic 
observations. Second, it minimizes the introduction of biases or noise that can arise from relying on multiple empirical relations or calibrations across different passbands. Third, it allows for more consistent and homogeneous metallicity estimates across diverse datasets, enhancing our ability to trace stellar populations of different metallicity in the MW and Local Group galaxies.

Recent studies have explored ML/DL approaches to predict RRL metallicities, either through regression on Fourier parameters \citep{Hajdu2018, Dekany2021, muraveva2025metallicity} or by leveraging the full light-curve information using architectures such as Recurrent Neural Networks (RNNs), including Long Short-Term Memory (LSTM) and Gated Recurrent Units (GRU), as well as Convolutional Neural Networks (CNNs) \citep{Dekany2022, Monti2024}.
ML methods have also demonstrated remarkable success in capturing complex patterns in time-series data, such as the light curves of variable stars \citep{belokurov2003light, belokurov2004light, wozniak2004identifying, willemsen2007study, debosscher2007, mahabal2008, Richards2011}. Furthermore, techniques like transfer learning have been employed to extend well-calibrated models from one photometric band to others with minimal additional noise \citep{Dekany2022}.

The application of DL directly to light curves has become both crucial and timely with the advent of the ESA {\it Gaia} mission \citep{Gaia2016} Data Release 3 (DR3, \citealt{Gaia2023}), which, among other data products, includes a catalogue of 271,779 RRLs across the whole sky \citep{Clementini2023gaia}. This catalogue provides time-series photometry in the $G$, $G_{BP}$, and $G_{RP}$ bands, as well as pulsation parameters (periods, amplitudes) for all RRLs in the sample. However, photometric metallicities were made available for only 133,577 RRLs (49\% of the sample), for which the Fourier parameter $\phi_{31}$ was calculated. Thus, retrieving metallicities directly from the $G$-band light curves by means of DL method would not only reduce systematics introduced by intermediate calibrations, but also nearly double the number of RRLs with known photometric metallicities. 
This approach will become even more relevant with the advent of {\it Gaia} Data Release 4 (DR4), currently expected in the second half of 2026, which will include time-series photometry for two billion stars.

In \citet{Monti2024}, we present a DL algorithm for estimating the photometric metallicities from the light curves of RRab stars. In this paper, we propose a unified DL approach for estimating the photometric metallicity of both RRab and RRc stars using {\it Gaia} DR3 $G$-band light curves. 
By leveraging advanced DL techniques, such as CNNs and RNNs, we aim to provide accurate metallicity predictions. 

The paper is organized as follows: Section~\ref{sec:data} describes the dataset and preprocessing steps. Section~\ref{sec:modeling} outlines the architecture and training process of the DL models. We also present strategies for model selection and optimization, along with performance results, including evaluations of the model’s accuracy. In Section~\ref{sec:validation}, we validate the derived metallicities. Finally, Section~\ref{sec:summary} summarizes our findings and discusses directions for future research.

\section{Photometric Data and Preprocessing}
\label{sec:data}

This study frames the prediction of photometric metallicity from light curves as a \textit{Time-series Extrinsic Regression} (TSER) problem, following the definition by \citet{Tan2021} and as applied in \citet{Monti2024}. TSER aims to establish a mapping from an entire time series (the light curve) to a continuous scalar value (metallicity). Building upon the definition by \citet{Tan2021}, TSER specifically addresses the task of predicting a single, static, continuous scalar value (the 'extrinsic' property) based on the characteristics of an entire input time series. This contrasts fundamentally with Time-Series Forecasting (TSF), where the goal is typically to predict future values within the sequence itself. In our context, metallicity is an intrinsic, time-invariant property of RRLs.


Essentially, the TSER model must learn the complex, potentially non-linear function $f: \mathcal{T} \rightarrow \mathbb{R}$, where $\mathcal{T}$ is the space of possible light-curve sequences (after appropriate preprocessing and standardization) and $\mathbb{R}$ represents the continuous metallicity scale. This involves identifying and utilizing patterns across the entire time series relevant to the extrinsic target variable. Thus, framing our problem as TSER accurately reflects the objective: regressing an extrinsic scalar property ([Fe/H]) from the complete photometric time series representing the star's pulsation.

\subsection{Data Selection and Initial Cleaning}\label{sec:cleaning}


The {\it Gaia} DR3 catalogue includes 270,891 RRLs analyzed and confirmed by the Specific Object Study pipeline for Cepheids and RRLs (SOS Cep\&RRL; \citealt{Clementini2023gaia}). The catalogue provides, among other parameters, pulsation periods, epochs of maximum light, mean magnitudes, and Fourier decomposition parameters of the $G$-band light curves, along with time-series photometry in the $G$, $G_{BP}$, and $G_{RP}$ bands. In \citet{muraveva2025metallicity}, we cleaned the sample of RRLs from the {\it Gaia} DR3 catalogue and presented new relations between the RRL metallicities, their pulsation periods, and the Fourier decomposition parameters published in DR3. These relations were calibrated using the spectroscopic metallicities available in the literature \citep{Crestani2021, Liu2020}. A feature selection algorithm was employed to identify the most relevant features for metallicity estimation. To fit the relations, we adopted a Bayesian approach, accounting for parameter uncertainties and the intrinsic scatter in the data. As a result, photometric metallicity estimates were derived for 134,769 RRLs (114,468 RRab and 20,001 RRc) from the clean {\it Gaia} DR3 sample.

\begin{figure*}
\centering
\includegraphics[width=0.7\hsize]{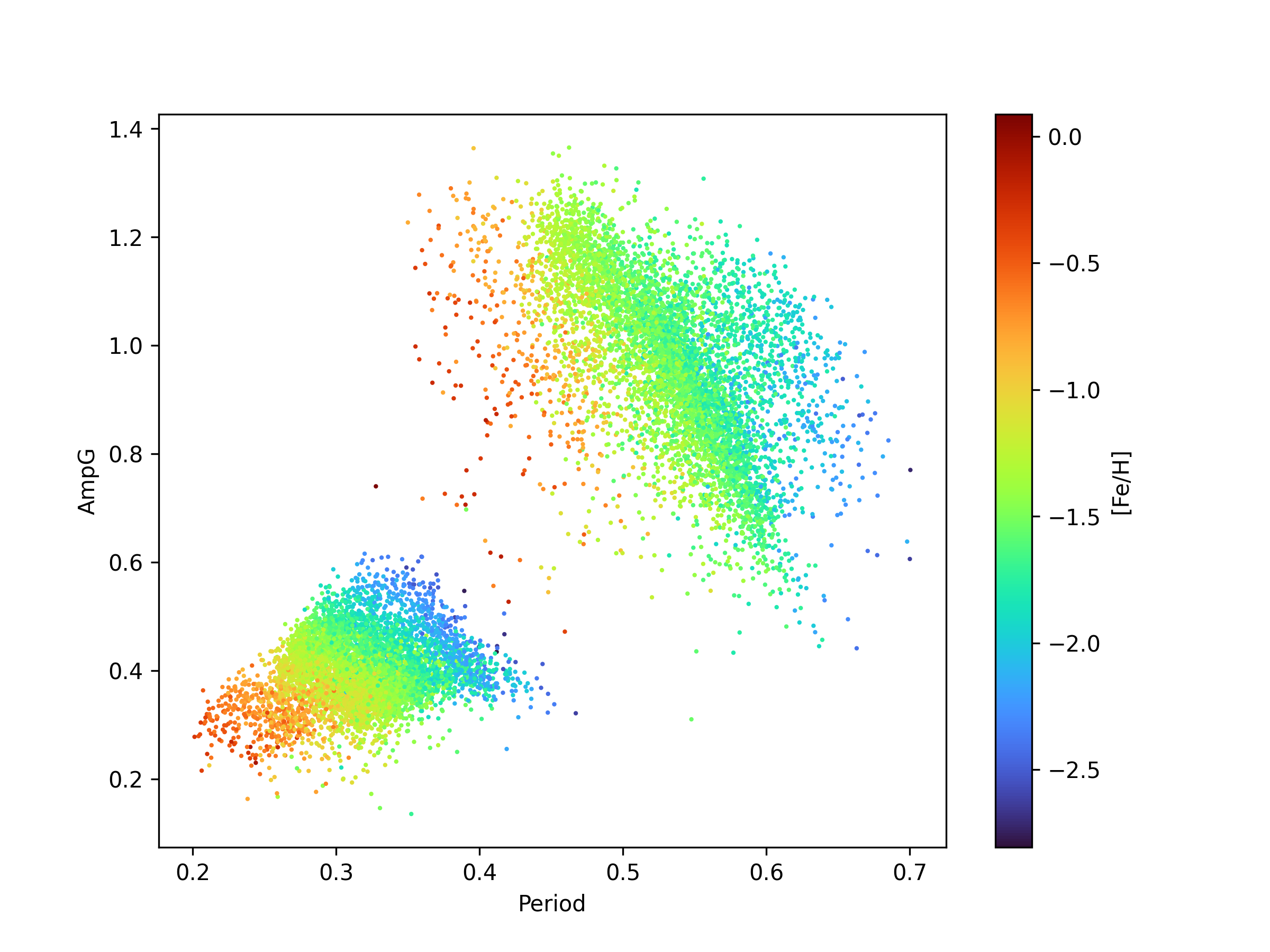}
    \caption{Distributions of RRab and RRc stars from our selected development datasets on the amplitude in the $G$ band versus period diagram, color-coded by metallicity.}
     \label{fig:per_amp}
\end{figure*}

In this study, we use the time-series photometry of RRLs in the $G$-band from the {\it Gaia} archive\footnote{\url{https://gea.esac.esa.int/archive/}}, along with pulsation periods from the \texttt{vari\_rrlyrae} table \citep{Clementini2023gaia}, for a cleaned sample of RRLs with available photometric metallicities from \citet{muraveva2025metallicity}. To ensure a high-quality dataset suitable for DL models, we applied stringent selection criteria to the RRL sample, targeting both RRab and RRc stars:

\begin{itemize}
    \item The uncertainty in the photometric metallicity estimate from \citet{muraveva2025metallicity} ($\sigma[\mathrm{Fe/H}]$) was constrained to be $\leq 0.4$ dex, ensuring reliable metallicity labels.
    \item The peak-to-peak amplitude in the {\it Gaia} $G$-band ($AmpG$) was limited to $\leq 1.4$ mag to exclude potential outliers or misclassified variables with unusually large amplitudes.
    \item Each light curve required a minimum of 50 observed epochs ($N_\mathrm{epochs} \geq 50$) in the $G$-band to ensure adequate phase coverage for reliable characterization.
    \item The uncertainty in the Fourier phase parameter $\phi_{31}$ ($\sigma\phi_{31}$) was required to be $\leq 0.10$ (only for RRab stars).
\end{itemize}
These criteria filter out sources with low-quality metallicity estimates or poorly sampled/characterized light curves, yielding a robust dataset. The final datasets satisfying these criteria consisted of 6002 RRab stars (4801 for training, 1201 for validation) and 6613 RRc stars (5290 for training, 1323 for validation). An illustrative example of the data structure is provided in Table \ref{tab:dataset_example}.
Figure~\ref{fig:per_amp} displays the amplitude in the $G$ band versus period diagram (Bailey diagram) for the selected RRab and RRc development datasets, color-coded by their [Fe/H] values. The diagram clearly illustrates the well-known dependence of amplitude on period and metallicity for RRL stars (e.g., \citealt{Clementini2023gaia}).

\subsection{Phase Folding and Alignment}
\label{subsec:phase_fold_align}

A fundamental step in analyzing periodic variable stars is phase folding. The phase ($\Phi$) for each observational transit was calculated using the following equation:

\begin{equation}
    \Phi = \left( \frac{T - \text{Epoch}_{\max}}{P} \right) - mod \left ( \frac{T - \text{Epoch}_{\max}}{P} \right ),
\end{equation}
where $T$ represents the observation time (that is: the Heliocentric Julian Day of observation - HJD), and $\text{Epoch}_{\max}$ and $P$ correspond to the epoch of maximum light and the pulsation period, respectively, as provided in the {\it Gaia} DR3 \texttt{vari\_rrlyrae} table \citep{Clementini2023gaia}. This process aligned the light curves to a common phase reference, facilitating direct comparisons across multiple cycles of variability. Aligning the light curves to the phase of maximum light is crucial, especially for the asymmetric RRab stars, ensuring consistent feature representation for the subsequent modeling steps (see e.g., \citealt{Dekany2022} for detailed discussion on phase alignment).


\subsection{Smoothing Spline Interpolation and Standardization}
\label{subsec:smoothing}
{\it Gaia} DR3 light curves often have uneven sampling and varying numbers of data points in the $G$-band, ranging from 51 to 256 in our development datasets.
To create uniform input sequences suitable for DL models and to reduce observational noise, a smoothing spline interpolation was applied to each phase-folded light curve using the SciPy library's \texttt{UnivariateSpline} function \citep{2020SciPy}. In more detail, this technique is used for fitting a smooth curve to a dataset, striking a balance between accurately representing the data and minimizing noise or fluctuations while standardizing the number of data points across all light curves. In this context, the data consists of light curves characterized by magnitude and phase, with each light curve containing a varying number of data points. The method involves determining a function that passes near the data points while minimizing the overall roughness or curvature of the curve. This approach is especially valuable when the data contains random noise, as it provides a clearer representation of underlying trends or patterns.

The smoothing spline achieves this by minimizing the sum of squared deviations between the fitted curve and the data points, while imposing a penalty on the curve's curvature. Mathematically, the problem is formulated as:

\begin{equation} \label{eq:spline} 
    \sum_{i=1}^{n}(y_i - f(x_i))^2 + \lambda \int (f''(x))^2 dx 
\end{equation}

Here, $f(x)$ is the smooth function being fitted, $(x_i,y_i)$ are the data points, $\lambda$ is a smoothing parameter that controls the trade-off between adherence to the data and smoothness of the curve, and $\int (f''(x))^2 dx$ penalizes excessive curvature by integrating the squared second derivative of the function. The result is a continuous representation of the light curve, from which we sampled a fixed number of points (264 points for RRab and 265 points for RRc stars, uniformly distributed in phase) to standardize the length of all input sequences. This process effectively reduces noise while preserving the underlying shape of the light curve. 

\begin{figure*}
    \begin{subfigure}[t]{.5\textwidth}
      \centering
      \includegraphics[width=1\linewidth]{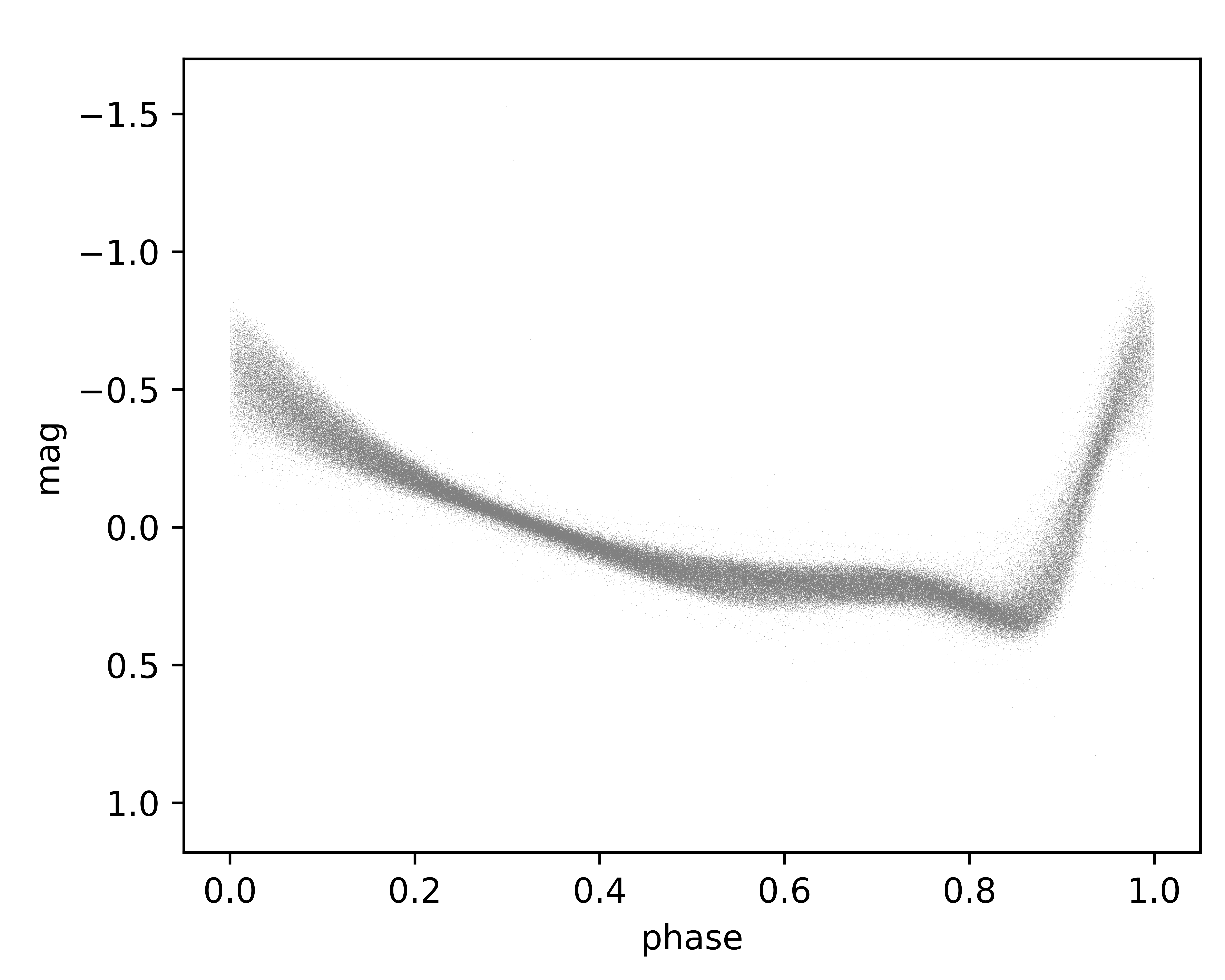}
      \caption{}
      \label{fig:phase_mag_ab}
    \end{subfigure}%
    \begin{subfigure}[t]{.5\textwidth}
      \centering
      \includegraphics[width=1\linewidth]{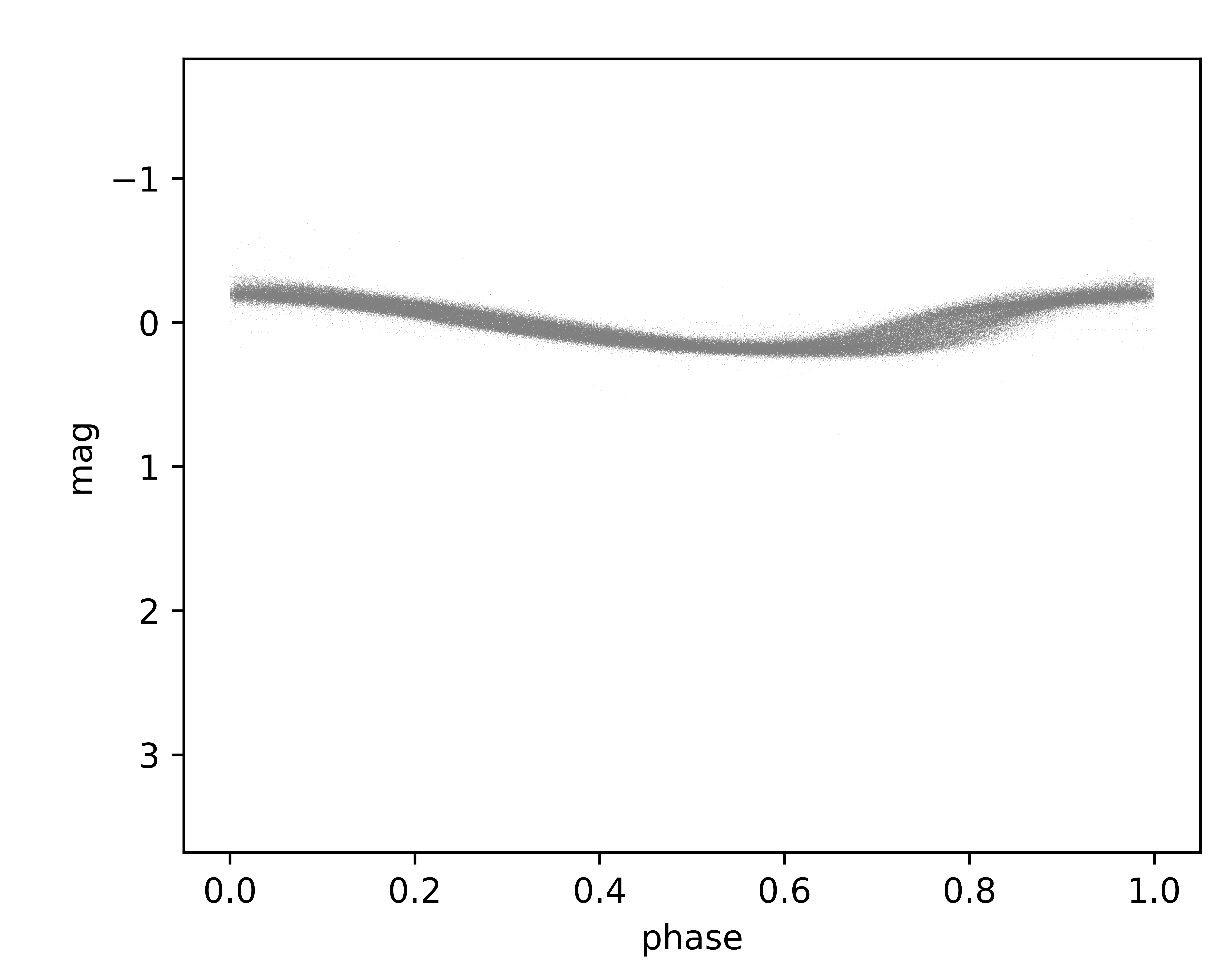}
      \caption{}
      \label{fig:phase_mag_c}
    \end{subfigure}
    \caption{Normalized splined $G$-band light curves of 6002 RRab stars (a) and 6613 RRc stars (b) from our development datasets.}
    \label{fig:phase_mag}
\end{figure*}

Following the spline interpolation and resampling, the magnitudes of each light curve were standardized. This crucial step was performed using the \texttt{StandardScaler} from the Scikit-learn library\footnote{\url{https://scikit-learn.org}} \cite{Pedregosa2011scikit}. For each light curve, this process involves subtracting its mean magnitude ($\mu_m$) and then dividing by its standard deviation of magnitudes ($\sigma_m$). From an astrophysical perspective, subtracting the mean magnitude effectively removes the star's average apparent magnitude over its pulsation cycle, thus centering the light curve around zero. This allows the subsequent analysis to be independent of the star's intrinsic luminosity, distance, and interstellar extinction, contributing to the observed mean magnitude. Subsequently, dividing by the magnitudes' standard deviation normalizes the light variation's amplitude. This ensures that all light curves are compared on a similar scale of variability, making the model focus on the \textit{shape} and \textit{morphological characteristics} of the light curve (e.g., skewness, acuteness of maxima/minima) rather than the absolute amplitude of pulsation, which can vary significantly even within the same class of variable stars. This standardization yields light curves with a mean of zero and a unit standard deviation, making them more suitable for training DL models by preventing features with larger numerical ranges from dominating the learning process. The resulting standardized light curves of RRab and RRc from our developement datasets are shown in Fig.\ref{fig:phase_mag}. 

\subsection{Sample Weights for Imbalanced Data}
\label{subsec:weights}
\begin{figure*}
    \begin{subfigure}[t]{.5\textwidth}
      \centering
      \includegraphics[width=1\linewidth]{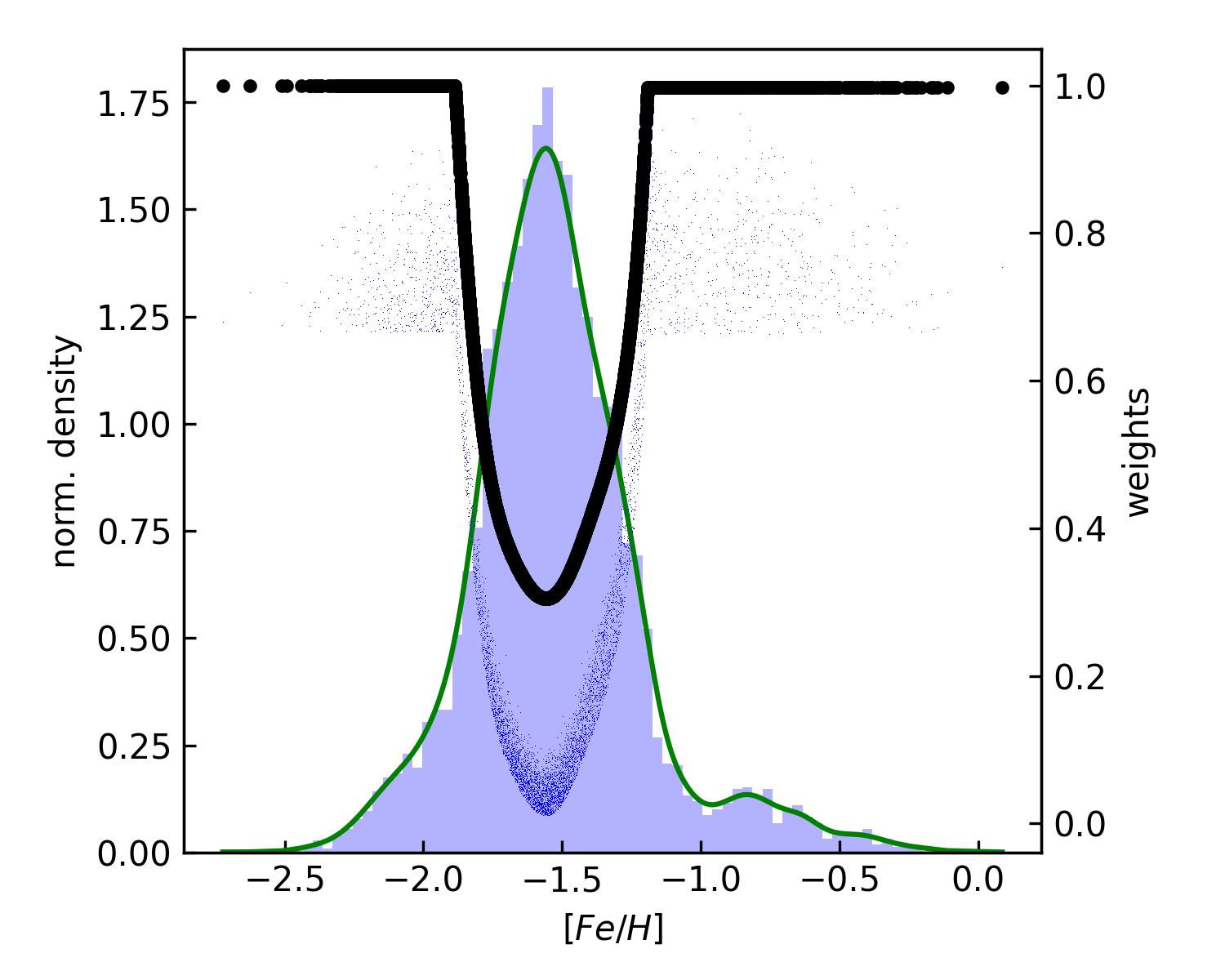}
      \caption{}
      \label{fig:metallicity_weights_ab}
    \end{subfigure}%
    \begin{subfigure}[t]{.5\textwidth}
      \centering
      \includegraphics[width=1\linewidth]{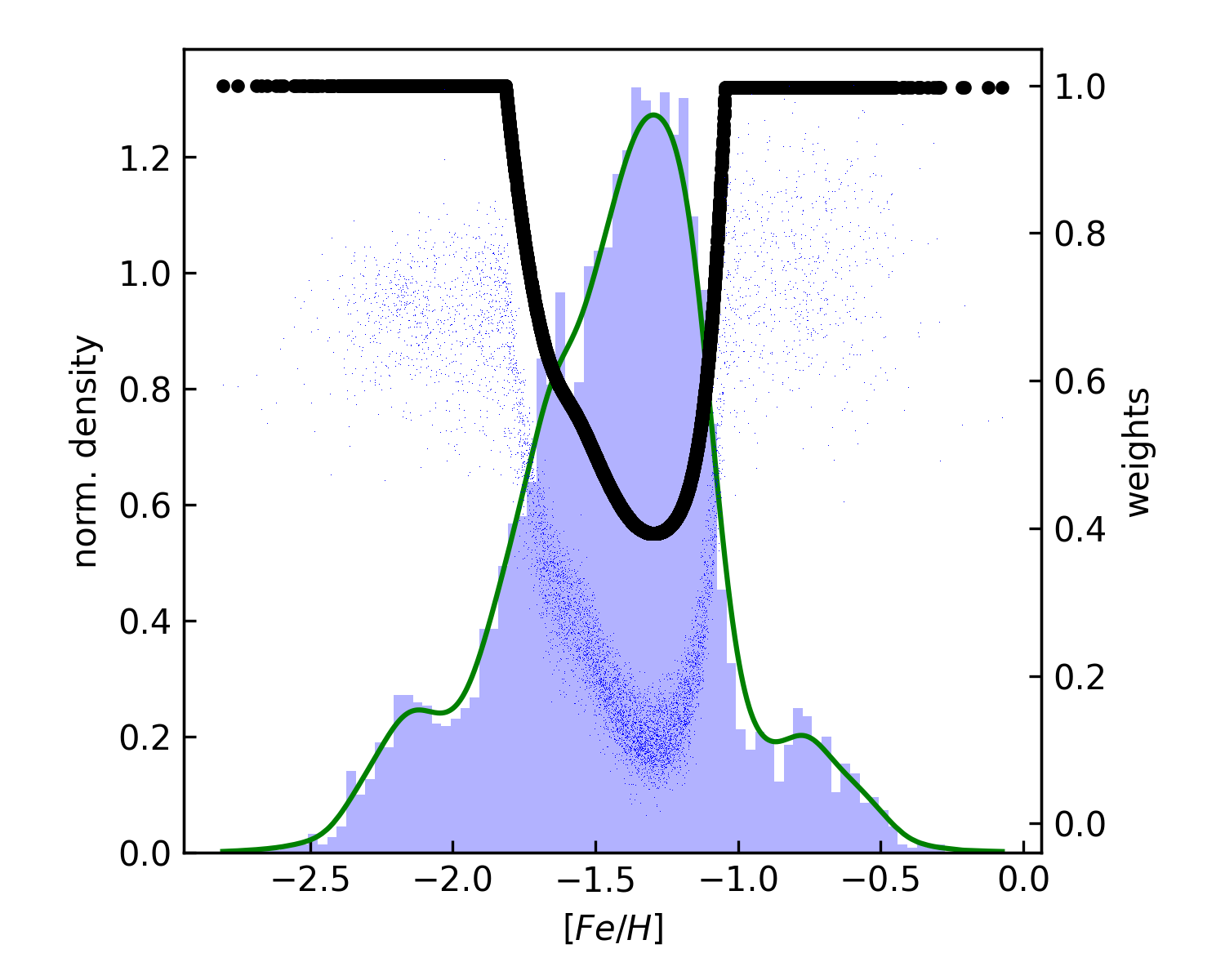}
      \caption{}
      \label{fig:metallicity_weights_c}
    \end{subfigure}
    \caption{Photometric metallicity distributions and corresponding sample weights (black lines) for RRab ({\it left panel}) and RRc ({\it right panel}) stars. Regions with lower data density are assigned higher weights to ensure a balanced contribution during model training.}
    \label{fig:metallicity_weights}
\end{figure*}

The photometric metallicity distribution of the RRLs in our development datasets is significantly unbalanced, with a pronounced peak around $[\mathrm{Fe/H}]\sim -1.5$~dex for RRab and $\sim-1.3$~dex for RRc stars (see Fig.~\ref{fig:metallicity_weights}). This imbalance can bias model training, as regions of the parameter space with fewer sources contribute less to the loss function during optimization. To address this issue, we introduce {\bf density-dependent sample weights} during training.

The sample weights ($w_d$) were computed using a Gaussian kernel density estimation (Gaussian KDE) approach:
\begin{equation}
    w_d = \frac{1}{\hat{\rho}(x)},
\end{equation}
where $\hat{\rho}(x)$ is the normalized density estimate of the metallicity distribution at a given point $x$. Gaussian kernels were used to estimate the density $\hat{\rho}$, ensuring that areas with lower data density received higher weights. These weights were normalized to ensure the sum of all weights equals the total number of samples, thereby preserving the overall loss scale.

The sample weights $w_d$ were integrated into the training pipeline as tensor inputs alongside the light curve data. Specifically:
\begin{itemize}
    \item The \textbf{input tensor} consists of the preprocessed light curve data: $$\mathbf{X} = \{X^{(t)}\}_{t=1}^{N_{\mathrm{ep}}}$$ where $X^{(t)}$ is defined as:
    \begin{equation}\label{eq:tensor_shape}
    X^{<t>} =
        \begin{cases} 
            m^{<t>} - <m> \\
            Ph \cdot P
        \end{cases}
        t = \{1,...,N_{ep}\}
    \end{equation}
    \item The \textbf{sample weights tensor} $$\mathbf{w} = \{w_d\}_{i=1}^N$$ was supplied as an additional input to the model's loss function. In practice, this means each sample in the dataset contributes to the loss function proportionally to its assigned weight $w_d$.
\end{itemize}

Integrating sample weights allows the model to account for the entire range of metallicities, mitigating bias caused by overrepresented regions in the dataset. Fig.~\ref{fig:metallicity_weights} illustrates the photometric metallicity distribution and the corresponding sample weights for RRab (left panel) and RRc (right panel) stars.

\subsection{Improved Model Performance with Uniform Preprocessing}
\label{subsec:model_performance}
As demonstrated by \cite{Monti2024}, preprocessing the phase-folded light curves significantly improves model performance. Applying noise reduction techniques, phase alignment, smoothing spline method, and sample weights enhances the predictive accuracy of DL models by ensuring consistent and high-quality input data. These improvements are particularly evident in regression tasks, where the preprocessing pipeline mitigates variability and biases inherent to raw observational data.

Both RRab and RRc stars were subjected to the identical preprocessing pipeline, ensuring the same degree of noise reduction, phase alignment, and normalization. Despite the intrinsic differences in their light curve shapes -- sawtooth-shaped for RRab and sinusoidal for RRc, clearly recognizable only in the optical bands, where the pulsation signatures are most prominent -- this unified approach guarantees comparable inputs for subsequent modeling, preserving the integrity of their physical and observational characteristics.

\section{Predictive modeling}
\label{sec:modeling}
\subsection{RNNs}
\label{subsec:rnn}
RNNs are a class of artificial neural networks designed to model sequential data by leveraging temporal dynamics. Unlike traditional feedforward neural networks, RNNs incorporate recurrent connections that enable them to maintain a hidden state, capturing information from previous time steps \citep{Williams1989}. This capability makes RNNs well-suited for tasks where the order and context of data points are critical, such as time series analysis \citep{Connor1994}, natural language processing \citep{Rodriguez1999}, and speech recognition \citep{Robinson1996}.

Mathematically, the hidden state \( h_t \) of an RNN at time \( t \) is updated as:
\begin{align}
    h_t = f(W_h \cdot h_{t-1} + W_x \cdot x_t + b),
\end{align}
where \( x_t \) is the input at time \( t \), \( W_h \) and \( W_x \) are weight matrices, \( b \) is the bias, and \( f \) is a non-linear activation function, typically \( \tanh \) or ReLU. The output of the RNN can either be a single value (for tasks like regression or classification) or a sequence (for tasks like translation or text generation). However, traditional RNNs face challenges such as the vanishing and exploding gradient problems, which limit their ability to learn long-term dependencies in sequences. To address these issues, advanced architectures like LSTM and GRU have been developed. These architectures introduce gating mechanisms that allow the network to selectively store, forget, or update information across time steps.

\subsection{GRU Neural Networks}
\label{subsec:gru}
The GRU, introduced by \cite{cho2014gru}, is a variant of the RNN designed to address the vanishing gradient problem in sequential data modeling. GRUs, like LSTM networks, capture long-term dependencies but are computationally lighter due to their simplified gating mechanism. The architecture of a GRU incorporates two primary gates: the \textit{update gate} and the \textit{reset gate}. 
The update gate \( z_t \) determines the amount of information from the past to retain, while the reset gate \( r_t \) controls the degree of forgetting of previous states. The mathematical formulation for GRU is as follows:
\begin{align}
    z_t &= \sigma(W_z \cdot [h_{t-1}, x_t] + b_z), \\
    r_t &= \sigma(W_r \cdot [h_{t-1}, x_t] + b_r), \\
    \tilde{h}_t &= \tanh(W_h \cdot [r_t * h_{t-1}, x_t] + b_h), \\
    h_t &= (1 - z_t) * h_{t-1} + z_t * \tilde{h}_t,
\end{align}
where \( x_t \) is the input, \( h_{t-1} \) is the hidden state from the previous timestep, \( W \) and \( b \) are learnable weights and biases, \( \sigma \) is the sigmoid activation function, and \( \tanh \) is the hyperbolic tangent activation function.

GRUs are particularly effective for regression tasks involving sequential data, such as time series prediction. To adapt GRUs for regression, the network is typically structured with a dense output layer having a linear activation function:
\begin{align}
    \hat{y} = W_o \cdot h_t + b_o,
\end{align}
where \( W_o \) and \( b_o \) are the weights and biases of the output layer, and \( \hat{y} \) is the predicted continuous value.

The network is trained to minimize a loss function suitable for regression, such as Mean Squared Error (MSE):
\begin{align}
    \mathcal{L}_{\text{MSE}} = \frac{1}{n} \sum_{i=1}^{n} (y_i - \hat{y}_i)^2,
\end{align}
where \( y_i \) and \( \hat{y}_i \) are the ground truth and predicted values, respectively.

The choice of GRU is further supported by our previous findings \cite{Monti2024}, where nine different sequential models were evaluated, and the GRU architecture consistently yielded the best performance.

\begin{figure*}
\centering
\includegraphics[width=0.68\hsize]{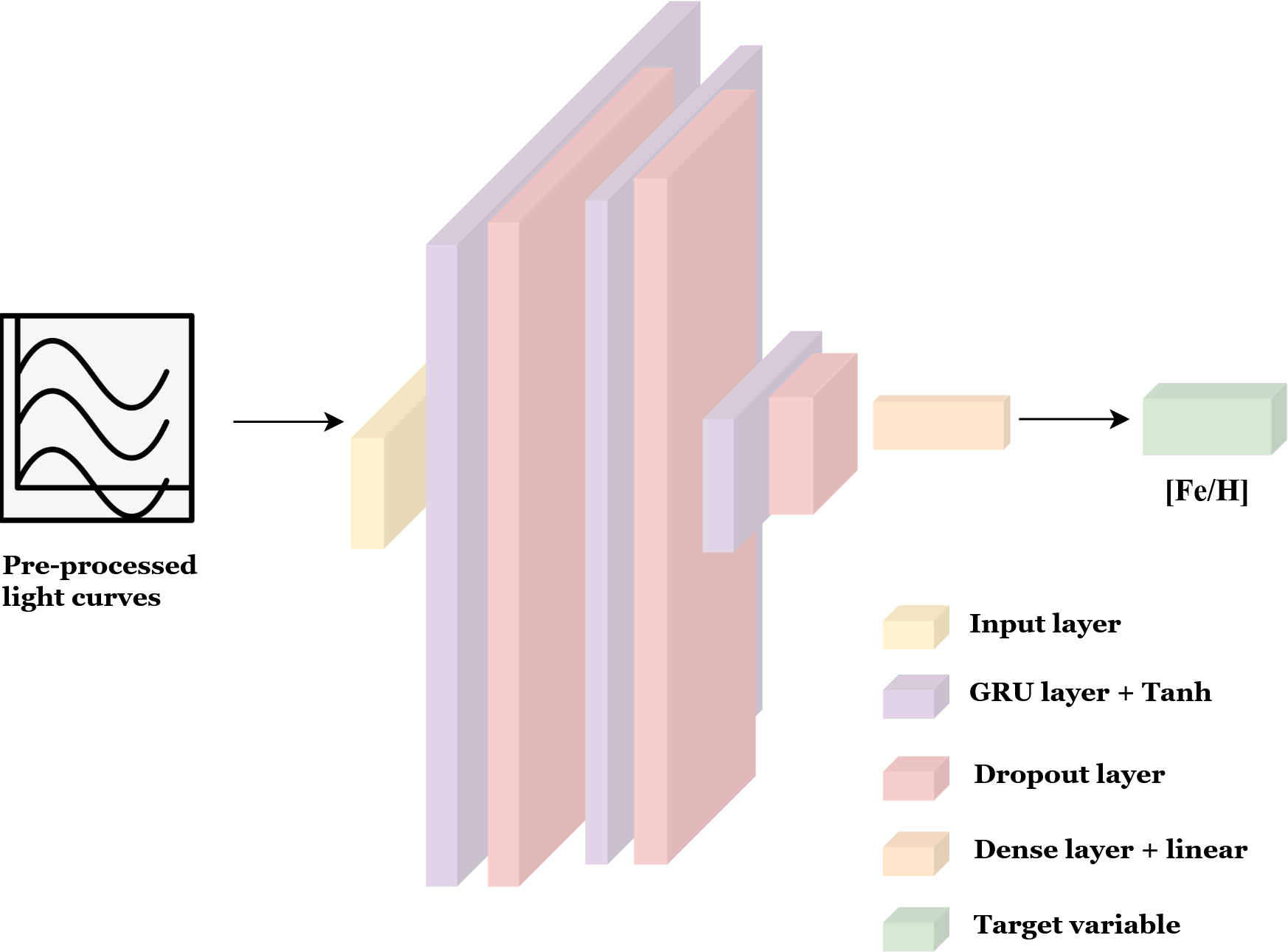}
    \caption{Schematic overview of the GRU-based neural network architecture used for predicting stellar metallicity ([Fe/H]) from pre-processed light curves. The model comprises an input layer, a sequence of GRU layers with Tanh activations, interleaved with dropout layers to prevent overfitting, followed by a dense linear layer producing the final regression output.}
     \label{fig:gru_architecture}
\end{figure*}

In our architecture shown in Fig.~\ref{fig:gru_architecture}, each main block comprises a \textit{GRU layer} followed by a \textit{dropout layer}, mitigating overfitting and enhancing the model's generalization capabilities. This approach is pivotal in improving the performance and robustness of models handling sequential data. The final configuration of the \textit{GRU} network includes three such main blocks, followed by a \textit{dense} layer with linear activation, as detailed in the formal description below:

\begin{equation}
    \begin{split}
        h_{t}^{(0)} &= X \\
        \textit{\text{for }} i &= 1 \textit{\text{ to }} n \\
        & h_{t}^{(i)} = \text{GRU\_block}(d_{t}^{(i-1)}) \\
        & d_{t}^{(i)} = \text{Dropout}(h_{t}^{(i)}) \\
        \textit{\text{end}}& \textit{\text{ for}}\\
        \hat{y} = &\text{Dense}(d_{t}^{(n)})
    \end{split}
\end{equation}

Here, $n$ represents the number of main blocks, set to 3 in our design, and $GRU\_block$ refers to a block constructed using a \textit{GRU layer}.

\subsection{Strategies for Model Selection and Optimization}
\label{subsec:model_selection}
The development of an effective predictive model involves two fundamental stages: \textit{training} and \textit{hyperparameter optimization}. During the training phase, the model's internal parameters (weights and biases) are adjusted iteratively to minimize a chosen loss function on a dedicated training dataset. For neural networks, this minimization is typically achieved using gradient-based optimization algorithms \citep{Goodfellow2016}. Concurrently, hyperparameter optimization focuses on selecting the best configuration for the model architecture and the training process itself. Hyperparameters — such as the number of layers, the number of neurons per layer, the learning rate, dropout probability, and the type of regularization — are not learned directly from the training data but are set beforehand. Their optimal values are determined by evaluating the model's performance on a separate validation dataset, ensuring the chosen configuration generalizes well to unseen data.

For efficient hyperparameter exploration across the nine different network architectures considered, we employed the \textit{Hyperband} algorithm \citep{Li2018hyperband}, as implemented in the Scikit-learn library \citep{Pedregosa2011scikit}. The search space included dropout rates in [0.1, 0.2, 0.4, 0.6], learning rates in [0.001, 0.01, 0.1], and batch sizes in [32, 64, 128, 256, 512].

The core of the \textit{training} process utilized the MSE as the loss function. Crucially, this loss was weighted using the sample weights derived in Section \ref{subsec:weights} to counteract the metallicity imbalance inherent in the RRL dataset (Fig.~\ref{fig:metallicity_weights}). To mitigate the risk of overfitting—where the model learns the training data too well, including its noise, and performs poorly on new data—we incorporated standard regularization techniques. Specifically, we experimented with \textit{kernel regularization} [L1 and L2 penalties, also known as Lasso \citep{Tibshirani1996} and Ridge \citep{Hoerl1970}, applied to network weights] and \textit{Dropout} \citep{Srivastava2014dropout}, which randomly sets a fraction of neuron outputs to zero during training, preventing over-reliance on specific features. The Hyperband search helped identify the optimal combination of these regularization strategies and their associated parameters (e.g., dropout rate, L1/L2 strength) for each architecture.

Evaluating model performance during \textit{hyperparameter tuning} and for final assessment requires metrics that accurately reflect prediction quality on unseen data. We used standard regression metrics: Root Mean Squared Error (RMSE) and Mean Absolute Error (MAE), along with their weighted counterparts (wRMSE and wMAE) that incorporate the sample weights. These metrics quantify the average prediction error magnitude. Additionally, we considered the coefficient of determination ($R^2$) score:

\begin{equation}
   R^2 = 1 - \frac{\sum_i (y_i - \hat{y}_i)^2}{\sum_i (y_i - \bar{y})^2},
   \label{eq:r2}
\end{equation}

where \( y_i \) represents the observed value of the dependent variable for the i-th observation, \( \hat{y}_i \) represents the value of the dependent variable predicted by the model for the i-th observation, and \( \bar{y} \) represents the mean of the observed values. The $R^2$ value ranges from 0 to 1, with 1 indicating perfect prediction and 0 indicating that it does not explain any variability in the dependent variable. Higher $R^2$ values suggest a better model fit.

To obtain robust estimates of model performance and generalization ability, we employed \textit{Repeated Stratified K-Fold Cross-Validation}. This technique extends standard K-Fold cross-validation. The dataset is divided into K folds (partitions). In each iteration, K-1 folds are used for training, and the remaining fold is used for validation. Stratification ensures that the distribution of metallicity values (the target value) is approximately preserved in each fold, which is crucial given the imbalanced nature of our data (Section \ref{subsec:weights}). The entire K-fold process is repeated multiple times with different random shuffles of the data before splitting into folds. This repetition reduces the dependence of the performance estimate on the specific random splits, yielding a more reliable assessment of how the model is likely to perform on new, unseen data.

For the actual parameter updates during training, we utilized the \textit{Adam} optimization algorithm \citep{Kingma2014adam}, a widely-used adaptive learning rate method known for its efficiency and effectiveness in DL tasks. A fixed learning rate of 0.01 was used, selected via the hyperparameter search. \textit{Early stopping} was employed as an additional regularization measure: training was halted when performance on the validation fold ceased to improve for a predefined number of epochs, preventing overfitting by stopping before the model starts fitting noise in the training data. A mini-batch size of 256 was used to balance computational efficiency with the stochasticity needed for effective gradient descent, while ensuring that each batch provided a reasonably comprehensive representation of the metallicity range. All code developed for this study is publicly available in the open-source GitHub repository\footnote{https://github.com/LorenzoMonti/metallicity\_rrls}.

\begin{figure*}
    \begin{subfigure}[t]{.5\textwidth}
      \centering
      \includegraphics[width=1\linewidth]{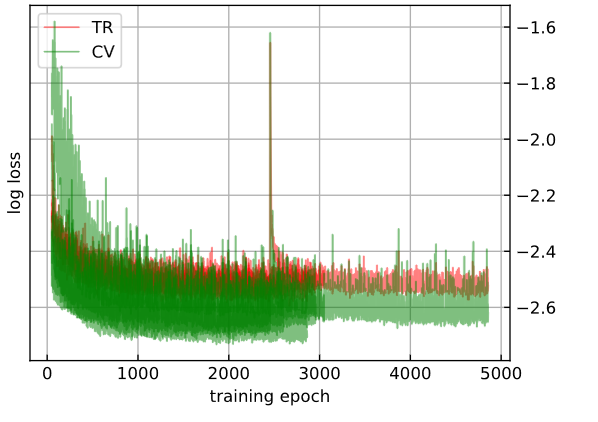}
      \caption{}
      \label{fig:progress_loss_ab}
    \end{subfigure}%
    \begin{subfigure}[t]{.5\textwidth}
      \centering
      \includegraphics[width=1\linewidth]{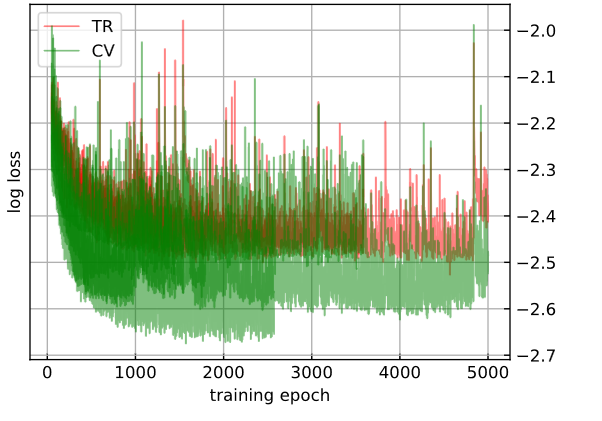}
      \caption{}
      \label{fig:progress_loss_c}
    \end{subfigure}
    \caption{
    Training (red) and validation (green) loss curves across epochs for each of the five cross-validation folds.The left panel (a) corresponds to RRab stars, while the right panel (b) represents RRc stars. A steady decrease in both losses indicates effective model learning, while the close alignment between training and validation loss across folds suggests good generalization and minimal overfitting. Darker colors denote greater consistency between folds.}
    \label{fig:progress_loss}
\end{figure*}

Fig.~\ref{fig:progress_loss} displays the learning curves for our best-performing model (\textit{GRU}, see Section \ref{subsec:gru}) during the \textit{Repeated Stratified K-Fold Cross-Validation} process. The plots show the evolution of the error's loss on both the training folds (red lines) and the corresponding validation fold (green lines) over the training epochs, separately for RRab (left panel) and RRc stars (right panel). For both stellar types, a clear and consistent decrease in both training and validation loss is observed as training progresses, indicating that the model is effectively learning the underlying relationship between the light curve features and metallicity. Importantly, the validation loss closely tracks the training loss without significant divergence, even across different folds (indicated by the overlap and darker colors where lines coincide). This behavior strongly suggests that the combination of our chosen model architecture (GRU), the preprocessing pipeline (Section \ref{sec:data}), and the regularization strategies employed (Dropout, L1/L2 penalties, and early stopping, as discussed above in Section \ref{subsec:model_selection}) successfully prevents overfitting. The model demonstrates good generalization capability, performing well not only on the data it was trained on but also on unseen data within the validation folds. The stability of the learning process across the multiple folds and repetitions further underscores the robustness of our training methodology and the resulting predictive model.

\subsection{Quantitative Performance Evaluation}
\label{subsec:evaluation}

Having established the model architecture (GRU), preprocessing pipeline (Section \ref{sec:data}), and optimization strategy (Section \ref{subsec:model_selection}), we now evaluate the quantitative performance of our final predictive model for TSER of RRL star's metallicity. The goal is to assess how accurately the model predicts the photometric [Fe/H] values derived by \citet{muraveva2025metallicity} based solely on the {\it Gaia} DR3 $G$-band light curves. The performance metrics, computed on the training and validation sets obtained through the \textit{Repeated Stratified K-Fold Cross-Validation} process described in Section \ref{subsec:model_selection}, are summarized in Table \ref{tab:gru_performance}. The results are presented separately for RRab and RRc stars.

Our optimized GRU model demonstrates high predictive accuracy and robust generalization. The coefficient of determination $R^2$ values are notably high: for RRab stars, $R^2 = 0.9447$ on the training set and $R^2 = 0.9401$ on the validation set. For RRc stars, the performance is even slightly better, with $R^2 = 0.9668$ (training) and $R^2 = 0.9625$ (validation). The close agreement between training and validation $R^2$ scores, along with the high absolute values, indicates that the model effectively captures the variance in metallicity explained by the light curve features and generalizes well to unseen data, avoiding significant overfitting.

The error metrics further corroborate the model's accuracy. The wRMSE, which accounts for the metallicity distribution imbalance (Section \ref{subsec:weights}), is low for both types: 0.0733 dex (RRab train), 0.0763 dex (RRab validation), 0.0679 dex (RRc train), and 0.0722 dex (RRc validation). Similarly, the wMAE values are also small, around 0.05-0.06 dex for both types and splits. The standard (unweighted) RMSE and MAE metrics show comparable low values, reinforcing the conclusion of high predictive precision across the board.

Compared to relevant prior work using DL on {\it Gaia} $G$-band data, specifically the BiLSTM model presented by \citet{Dekany2022}, our GRU model shows improved performance. For instance, the validation wRMSE achieved by our GRU model (0.0763 dex for RRab stars) is substantially lower than the 0.13 dex reported for the BiLSTM model by \citet{Dekany2022}, indicating a significant reduction in prediction error with our approach. Similar improvements are seen across other error metrics (wMAE, RMSE, MAE) when comparing Table \ref{tab:gru_performance} with the results in table~3 of \citet{Dekany2022}.

\begin{table}[h]
\caption{Results for RRab and RRc across various metrics for Training (Train) and Validation (Val) datasets using our final GRU model.}
    \centering
    \begin{tabular}{lcccccc}
    \toprule
    \multicolumn{2}{c}{} & \multicolumn{2}{c}{\textbf{RRab}} & \multicolumn{1}{c}{} & \multicolumn{2}{c}{\textbf{RRc}}\\ 
    \midrule
    Metrics & & Train & Val & & Train & Val \\
    \midrule
    $R^2$ & & 0.9447 & 0.9401 & & 0.9668 & 0.9625 \\
    wRMSE & & 0.0733 & 0.0763 & & 0.0679 & 0.0722 \\
    wMAE  & & 0.0547 & 0.0563 & & 0.0490 & 0.0504 \\
    RMSE  & & 0.0735 & 0.0765 & & 0.0681 & 0.0720 \\
    MAE   & & 0.0549 & 0.0565 & & 0.0492 & 0.0505 \\
    \bottomrule
    \end{tabular}
    \label{tab:gru_performance}
\end{table}

\begin{figure*}
    \begin{subfigure}[t]{.5\textwidth}
      \centering
      \includegraphics[width=1\linewidth]{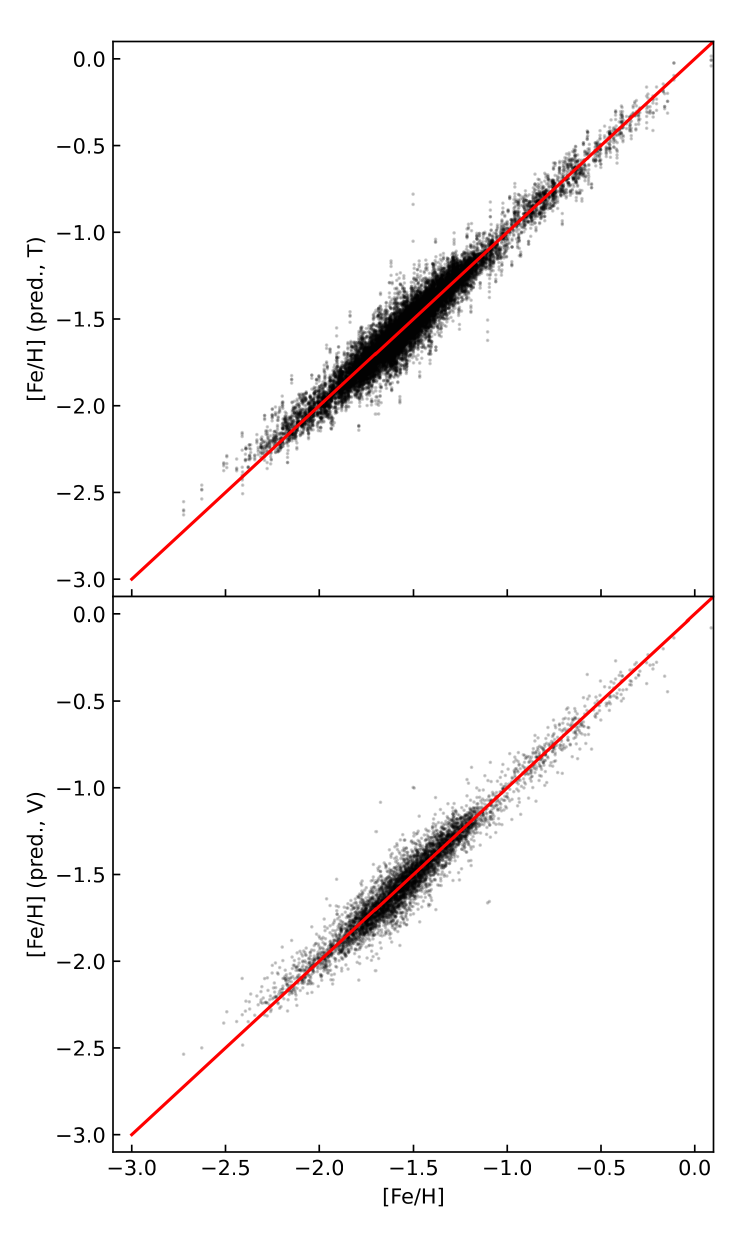}
      \caption{}
      \label{fig:pred_vs_true_ab}
    \end{subfigure}%
    \begin{subfigure}[t]{.5\textwidth}
      \centering
      \includegraphics[width=1\linewidth]{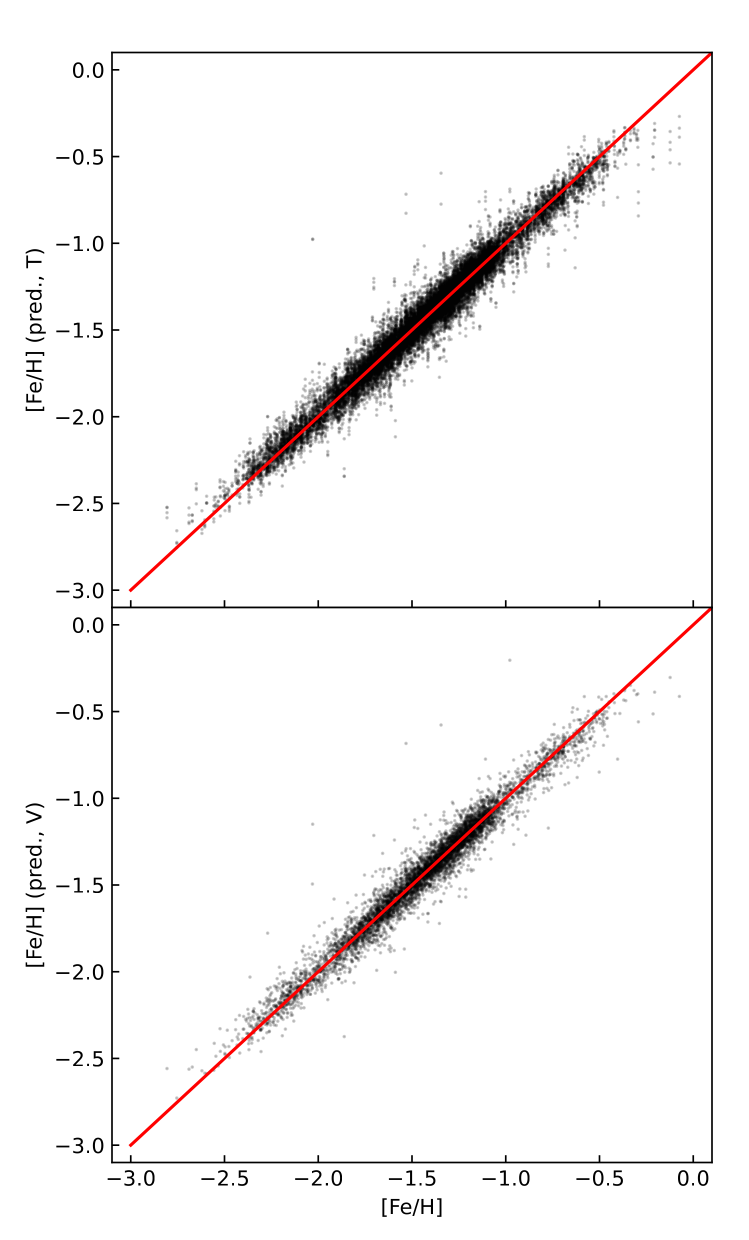}
      \caption{}
      \label{fig:pred_vs_true_c}
    \end{subfigure}
    \caption{
 True versus predicted photometric metallicity values from the \textit{GRU} predictive model for RRab ({\it left panel}) and RRc stars ({\it right panel}). In each case, the top and bottom panels correspond to the training (T) and validation (V) datasets, respectively. The red lines denote the identity function.}
    \label{fig:pred_vs_true}
\end{figure*}

Visual confirmation of the model's performance is provided in Fig.~\ref{fig:pred_vs_true}. These scatter plots compare the true (input) photometric metallicities against the values predicted by the GRU model. The panels show results for both RRab (left) and RRc (right) stars, separated into training (top) and validation (bottom) sets. In all cases, the data points cluster tightly around the identity line (y=x, shown in red), indicating excellent agreement between predicted and true values. The low amount of scatter visually confirms the low error metrics reported in Table \ref{tab:gru_performance} and reinforces the model's ability to accurately regress metallicity across the studied range.

Overall, the quantitative evaluation demonstrates that our optimized GRU-based model provides highly accurate and robust predictions of photometric metallicity for both RRab and RRc stars using {\it Gaia} DR3 $G$-band light curves, surpassing the performance of previously published DL models for this specific task and dataset.

\section{Validation of Photometric Metallicity Predictions}
\label{sec:validation}

Having developed and optimized our GRU-based DL model for the TSER task (Section \ref{subsec:model_selection}) and quantitatively assessed its performance using cross-validation (Section \ref{subsec:evaluation}), we now turn to a broader validation of its predictions. The goal is to understand how the model behaves across the large dataset of RRL stars for which target photometric metallicities from \citet{muraveva2025metallicity} are available, and to investigate factors influencing the prediction accuracy, such as pulsation type and observational sampling. 

The validation process involves applying the trained GRU model to infer [Fe/H] from the {\it Gaia} DR3 $G$-band light curves (preprocessed as described in Section \ref{sec:data}) and comparing these predictions  against the target photometric metallicity values  derived by \citet{muraveva2025metallicity}. It is important to recall that these target values, while serving as the ground truth for training and evaluating our specific regression model, are themselves photometric estimates calibrated using spectroscopically measured metallicities from the literature \citep{Crestani2021, Liu2020}. The comparison thus assesses how well our DL model, which utilizes the full time-series information, reproduces the metallicities derived from period-Fourier parameter relations.

\begin{figure*}[h]
\centering
\includegraphics[width=17cm, trim=40 10 10 0]{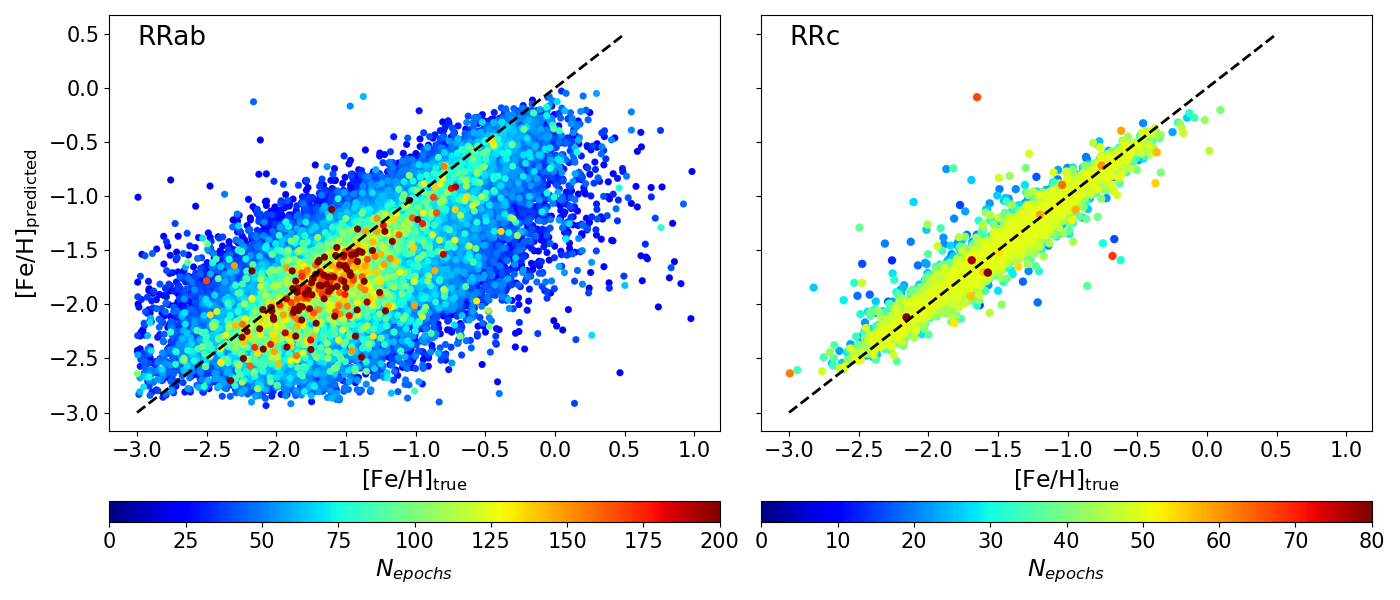}

    \caption{Comparison of true photometric metallicity values from \citet{muraveva2025metallicity} with the metallicity predicted by the GRU model for 108,766 RRab ({\it left panel}) and 13,388 RRc stars ({\it right panel}) in the validation datasets. The black dashed lines indicate the identity function (y=x). Each point is color-coded according to the number of $G$-band epochs ($N_\mathrm{epochs}$) available in the {\it Gaia} DR3 catalogue for each star.}\label{fig:comparison_all}
     
\end{figure*}

The left panel of Fig.~\ref{fig:comparison_all} presents this comparison for a substantial dataset comprising 108,766 RRab stars from the {\it Gaia} DR3 catalogue \citep{Clementini2023gaia}, which were not used in model training and for which photometric metallicities are provided by \citet{muraveva2025metallicity}. The scatter plot compares the model's predicted [Fe/H] against the true [Fe/H] values, with points color-coded according to the number of $G$-band epochs ($N_\mathrm{epochs}$) available in the {\it Gaia} DR3 catalogue for each star. A strong correlation around the identity line (y=x) is evident, confirming the model's general ability to predict RRab metallicities accurately, consistent with the high $R^2$ value reported in Table~\ref{tab:gru_performance}. However, a noticeable scatter is present, which clearly correlates with $N_\mathrm{epochs}$. Stars with fewer epochs (bluish colors) exhibit significantly larger deviations from the identity line compared to those with higher epoch counts (reddish colors). This trend strongly suggests that denser temporal sampling allows the GRU model to better learn the subtle morphological features of the light curve (e.g., rise time, asymmetry, presence of bumps) that correlate with metallicity. Sparse sampling inevitably leads to poorer phase coverage and increased uncertainty in the light curve representation, hindering the model's predictive precision. Moreover, it is important to stress that the RRab stars in the validation dataset were not included in the training set and may have larger errors in their metallicities or $\phi_{31}$ parameter, or a small number of epochs (according to the selection criteria described in Section~\ref{sec:cleaning}), and thus have less accurate photometric metallicities than the stars in the training set. These less accurate target photometric metallicities may have contributed to the scatter observed in the left panel of Fig.~\ref{fig:comparison_all}.
The inherent complexity and star-to-star variability in RRab light curve shapes --e.g., due to the Blazhko effect \citep{Blaszko1907}, although not explicitly modeled here-- may also contribute to the baseline scatter even at high epoch counts. 
Finally, the left panel of Fig.~\ref{fig:comparison_all} shows an offset between the predicted and true metallicity values for RRab stars, with the predicted values being systematically lower. However, the large overall scatter makes it difficult to determine the exact cause of this offset. A more thorough analysis will be possible once additional epochs become available for each star, with Gaia DR4 which spanning 66 months of observations (compared to the 34 months of DR3) will almost double the number of epochs, hence allowing to improve the reliability of both the GRU model predictions and the underlying photometric metallicity estimates.

The right panel of Fig.~\ref{fig:comparison_all} shows the analogous comparison for a large sample of 13,388 RRc stars. Visually, the correlation appears tighter, and the overall scatter around the identity line is reduced compared to the RRab sample, partly due to the significantly smaller number of stars for which this comparison was possible. The reduced scatter compared to RRab stars is consistent with the slightly better quantitative metrics obtained for RRc stars (Table~\ref{tab:gru_performance}). The improved performance for RRc stars is likely attributable to their light curve morphology. The light curves of RRc stars are more symmetric and closer to sinusoidal, and likely better sampled thanks to the shorter pulsation periods of RRc compared to RRab stars, hence presenting potentially simpler or more stable features for the model to correlate with [Fe/H]. While RRc stars can also exhibit modulation effects, their fundamental shape is less complex than that of RRab stars. This relative simplicity might make the metallicity inference less sensitive to sparse sampling or observational noise compared to the RRab stars case. Nonetheless, residual dependence on the number of epochs remains visible, although much less pronounced than for RRab stars, with higher $N_\mathrm{epochs}$ leading to more precise predictions. 

\begin{figure*}[h]
\centering
\includegraphics[width=17cm, trim=300 150 300 20]{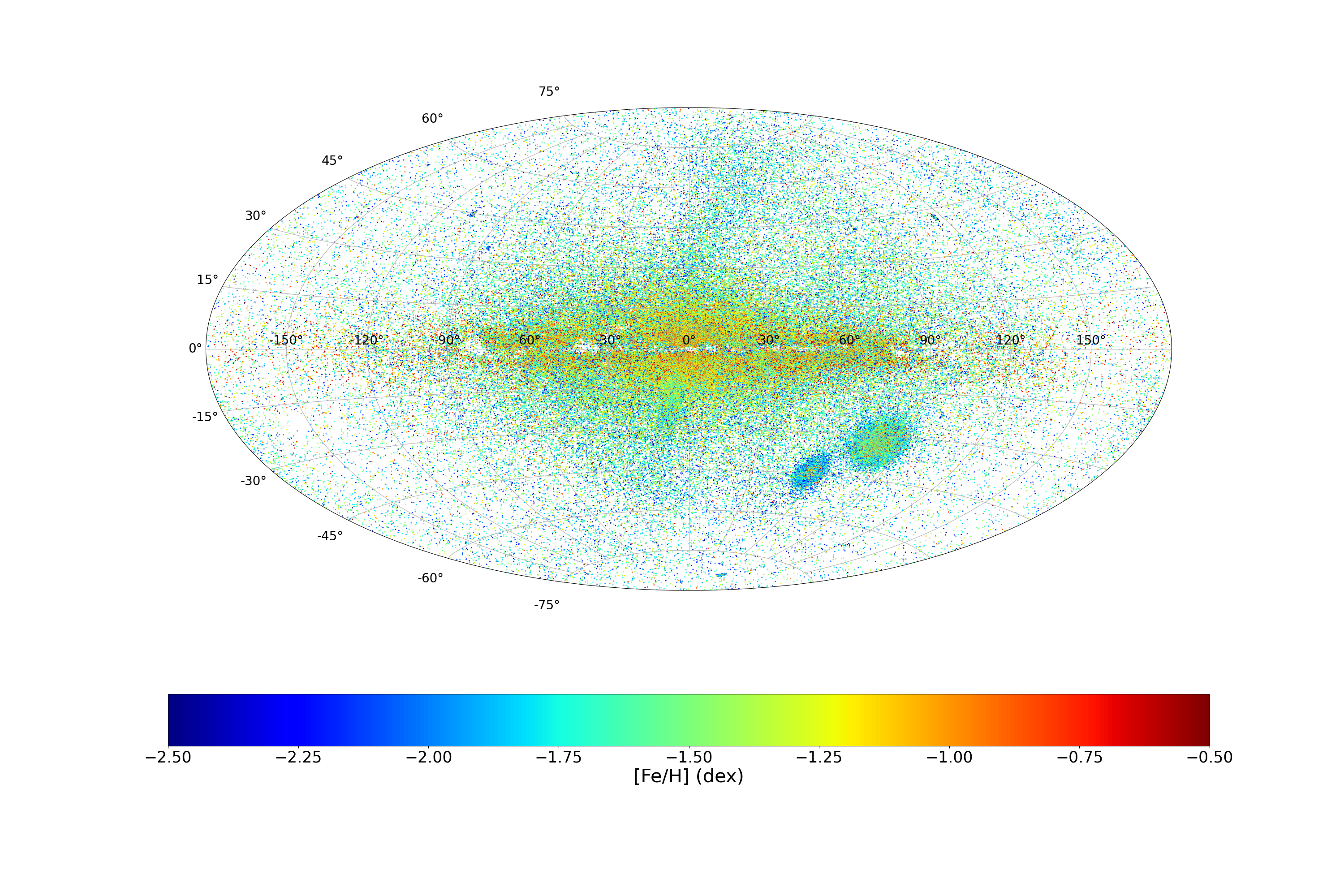}

    \caption{Sky distribution of 258,696 RRLs from the cleaned {\it Gaia} DR3 sample, color-coded by photometric metallicities derived using the GRU-based predictive model.}\label{fig:map}
     
\end{figure*}

Photometric metallicities calculated using period–Fourier parameters–metallicity relations were provided for 134,769 RRLs (114,768 RRab and 20,001 RRc) by \citet{muraveva2025metallicity}. The smaller number of RRc stars is due to the limited availability of the Fourier parameter $\phi_{31}$ for RRc stars in the {\it Gaia} DR3 catalogue, which is expected to significantly improve with DR4. By applying our DL model directly to the light curves, we were able to recover metallicities for 258,696 RRLs (169,024 RRab and 89,672 RRc stars) from the cleaned {\it Gaia} DR3 catalogue, increasing the number of stars with available metallicities by factors of 1.25 and 4.48 for RRab and RRc stars, respectively. This improvement has the potential to significantly enhance studies of the structure and chemical abundances of the MW and Local Group galaxies. Fig.~\ref{fig:map} shows the sky distribution of these 258,696 RRLs, color-coded by photometric metallicities derived by applying the DL model to their light curves. As expected, more metal-rich stars are concentrated in the Disk of the Galaxy, while more metal-poor stars are distributed throughout the MW halo. This demonstrates the potential of the method developed in this study for future applications.
 
The clear dependence of prediction accuracy on the number of observational epochs, demonstrated in Fig.~\ref{fig:comparison_all}, has significant implications. 
Based on the trends observed in Fig.~\ref{fig:comparison_all}, we anticipate that applying our GRU model (or similar DL approaches) to {\it Gaia} DR4 light curves will yield substantially more precise photometric metallicity estimates. The reduced scatter for high-$N_\mathrm{epochs}$ stars in the current data suggests that improved phase coverage directly translates to better constraints on metallicity-sensitive light curve features.

More precise and accurate large-scale photometric metallicity maps, derived from hundreds of thousands of RRLs distributed throughout the Galactic halo, bulge, and satellite systems, will enable unprecedented studies of the MW's chemical structure, accretion history, and the properties of its oldest stellar populations \citep{Dekany2022, Monti2024, muraveva2025metallicity}. Furthermore, future high-cadence, deep surveys like the Vera C. Rubin Observatory's Legacy Survey of Space and Time (LSST) will provide light curves with even better sampling for vast numbers of faint RRLs in the South emisphere. These future datasets promise to further refine TSER models, pushing the boundaries of precision achievable for photometric metallicity estimation and enabling detailed chemo-dynamical studies across enormous volumes of the Galaxy.

\section{Summary and Conclusions}
\label{sec:summary}

The current era of large-scale astronomical surveys, prominently featuring the {\it Gaia} mission, delivers datasets of unprecedented size and richness, necessitating the development of sophisticated computational tools for scientific analysis. RRLs stand out as essential probes of the MW's old, metal-poor populations. Estimating their metallicity is crucial for understanding Galactic chemical evolution, yet traditional methods to measure photometric metallicities face limitations in calibration and in capturing the nuances of light curve morphology. Addressing this, in our work we developed and validated a DL framework for the TSER of photometric metallicity, specifically applying a GRU architecture to {\it Gaia} DR3 $G$-band light curves of both RRab and RRc stars.

Our methodology integrated several critical components for robust results. A comprehensive data preparation pipeline  ensured data quality through careful selection, phase folding and alignment appropriate for pulsating stars, smoothing spline interpolation for noise reduction and standardization, and density-dependent sample weighting to counteract the natural metallicity imbalance of the Galactic RRL population. Rigorous hyperparameter optimization using Hyperband, combined with effective regularization strategies including L1/L2 penalties, Dropout, and early stopping within a Repeated Stratified K-Fold Cross-Validation framework, led to the selection of an optimized GRU model.

This final GRU model demonstrated high predictive accuracy and strong generalization capabilities. On the validation sets, it achieved high coefficients of determination ($R^2 \approx 0.94$ for RRab, $R^2 \approx 0.96$ for RRc) and low weighted root mean squared errors (wRMSE $\approx 0.076$ dex for RRab, wRMSE $\approx 0.072$ dex for RRc). These results represent a significant improvement in precision compared to previous DL benchmarks applied to {\it Gaia} $G$-band data \citep{Dekany2022}. Furthermore, our validation analysis explicitly quantified the positive impact of increased observational sampling; light curves with a higher number of epochs in the $G$-band consistently yielded more precise metallicity predictions. Finally, the application of our DL model directly to the light curves allowed us to increase the number of stars with available metallicities by factors of 1.25 and 4.48 for RRab and RRc stars, respectively.

The success of this GRU-based TSER approach highlights the broader potential of DL applied to modern astronomical time-series. Deriving accurate photometric metallicities directly from light curves offers a scalable and computationally efficient pathway for chemically characterizing vast numbers of RRLs, complementing or substituting more resource-intensive spectroscopic methods. This enables the construction of large, homogeneous metallicity catalogues directly linked to {\it Gaia}'s photometry and astrometry, facilitating detailed investigations into the structure, formation history, and chemical enrichment patterns of the MW and Local Group galaxies.

Looking forward, the demonstrated dependence on data sampling points towards substantial gains in precision with upcoming datasets. {\it Gaia} DR4, with its longer time baseline and an almost doubled number of epoch data compared to DR3 \citep{Gaia2023}, and the future deep, high-cadence observations from the Vera C. Rubin Observatory's LSST, will provide significantly richer light curves. Our future work aims to leverage these resources by expanding the scope of our models to potentially include more different photometric bands, further refining the DL architectures perhaps through attention mechanisms or Transformers, and ultimately applying the derived high-precision metallicity catalogues to pressing questions in Galactic archaeology and chemo-dynamics.

\begin{acknowledgements}
This work uses data from the European Space Agency mission {\it Gaia} (\url{https://www.cosmos.esa.int/gaia}), processed by the {\it Gaia} Data Processing and Analysis Consortium (DPAC; \url{https://www.cosmos.esa.int/web/gaia/dpac/consortium)}. Funding for the DPAC has been provided by national institutions, in particular the institutions participating in the {\it Gaia} Multilateral Agreement. Support to this study has been provided by INAF Mini-Grant (PI: Tatiana Muraveva), by the Agenzia Spaziale Italiana (ASI) through contract ASI 2018-24-HH.0 and its Addendum 2018-24-HH.1-2022, and by Premiale 2015, MIning The Cosmos - Big Data and Innovative Italian Technology for Frontiers Astrophysics and Cosmology (MITiC; P.I.B.Garilli). This research was also supported by the International Space Science Institute (ISSI) in Bern, through ISSI International Team project \#490, ‘SH0T: The Stellar Path to the H0 Tension in the {\it Gaia}, Transiting Exoplanet Survey Satellite (TESS), Large Synoptic Survey Telescope (LSST), and James Webb Space Telescope (JWST)
Era’ (PI: G. Clementini).
\end{acknowledgements}

%
%
\bibliographystyle{aa}
\bibliography{bib}

\begin{appendix} 
\clearpage
\section{Datasets}

\begin{table}[!htbp]
\caption{Parameters of 6002 RRab (a) and 6613 RRc (b) stars from the {\it Gaia} DR3 catalogue \citep{Clementini2023gaia} selected as discussed in Sect.\ref{sec:cleaning}.}
    \centering
    \bigskip
  
    \begin{tabular}{ccccccc}
    \multicolumn{7}{@{}l}{\em(a) RRab stars}\\\toprule
        \textbf{id} & \textbf{source\_id} & \textbf{period} & \textbf{AmpG} & \textbf{$N_\mathrm{epochs}$} & \textbf{{[}Fe/H{]}} & \textbf{$\sigma${[}Fe/H{]}}\\\midrule
        0           & 5978423987417346304 & 0.415071        & 0.61029154 & 53             & -0.144963           & 0.398111                    \\
        1           & 5358310424375618304 & 0.407642        & 0.6174223  & 56             & -0.223005           & 0.391468                    \\
        2           & 5341271082206872704 & 0.327778        & 0.7399841  & 53             &  0.087612           & 0.382031                    \\
        3           & 5844089608021904768 & 0.459576        & 0.47177884 & 54             & -0.380516           & 0.396500                    \\
        4           & 5992931321712867200 & 0.390948        & 0.76943225 & 63             & -0.256892           & 0.391830                    \\
        ...         & ...                 & ...             & ...        & ...            & ...                 & ...                         \\
        5997        & 5917421845281955584 & 0.532958        & 0.9153245  & 52             & -1.507490           & 0.379151                    \\
        5998        & 4659766188753815552 & 0.413777        & 0.9984105  & 245            & -0.758832           & 0.373709                    \\
        5999        & 5868263951719014528 & 0.365109        & 1.0959375  & 66             & -0.300124           & 0.380579                    \\
        6000        & 5963340573264428928 & 0.452752        & 1.0733474  & 64             & -1.079237           & 0.369655                    \\
        6001        & 5796804423258834560 & 0.510323        & 1.0356201  & 57             & -1.553741           & 0.372771                    \\\bottomrule
    \end{tabular}

    \bigskip
    \bigskip
    \begin{tabular}{ccccccc}
    \multicolumn{7}{@{}l}{\em(b) RRc stars}\\\toprule
        \textbf{id} & \textbf{source\_id} & \textbf{period} & \textbf{AmpG} & \textbf{{$N_\mathrm{epochs}$}} & \textbf{{[}Fe/H{]}} & \textbf{$\sigma${[}Fe/H{]}}\\\midrule
        0           & 5596822911942653184 & 0.229658        & 0.266563 & 64             & -0.510573           & 0.319436                    \\
        1           & 5979996357763478656 & 0.238414        & 0.345428 & 60             & -0.774683           & 0.308779                    \\
        2           & 5980315765840586368 & 0.239315        & 0.331957 & 58             & -0.784762           & 0.329183                    \\
        3           & 5962392935998706944 & 0.236403        & 0.339708 & 56             & -0.781534           & 0.329995                    \\
        4           & 6026953333884968576 & 0.225267        & 0.335460 & 66             & -0.660949           & 0.316708                    \\
        ...         & ...                 & ...             & ...        & ...            & ...                 & ...                         \\
        6608        & 5283973542734756608 & 0.234871        & 0.401754  & 189             & -0.862106           & 0.316825                    \\
        6609        & 4689034020082476416 & 0.230059        & 0.361965  & 52            & -0.706349             & 0.341561                    \\
        6610        & 5941537227647686528 & 0.315373        & 0.332867  & 59             & -1.114446            & 0.338075                    \\
        6611        & 4312957539614846336 & 0.284292        & 0.331146  & 54             & -1.076717            & 0.333702                    \\
        6612        & 5865806779419912320 & 0.393082        & 0.378917  & 73             & -1.981822            & 0.360913                    \\\bottomrule
    \end{tabular}
  \tablefoot{Column (1) identification number; Column (2) {\it Gaia} DR3 source\_id; Column (3) Pulsation period (days); Column (4) Amplitude in the $G$ band (mag); Column (5) Number of epochs in the $G$ band; Columns (6) and (7) Photometric metallicity and errors from \cite{muraveva2025metallicity}.}
    
\label{tab:dataset_example}
\end{table}

\subsubsection{Computational Environment}
\label{subsec:setup}
The training and evaluation of the DL models described in this work were computationally intensive, requiring specialized hardware. All experiments were conducted on a high-performance workstation equipped with an NVIDIA GeForce RTX 4070 graphics processing unit (GPU), featuring 12GB of VRAM, specifically chosen for its capability to accelerate the matrix operations inherent in neural network training. The system was further configured with 32GB of DDR5 RAM for efficient data handling and a high-speed NVMe solid-state drive (SSD) to minimize data loading bottlenecks during training.

The software environment was based on Python version 3.10. The core deep learning framework utilized was TensorFlow version 2.13.0, accessed via the Keras API version 2.13.1, for model definition, training, and evaluation. To leverage the GPU's computational power, the CUDA Toolkit version 11.8 and the NVIDIA CUDA Deep Neural Network library (cuDNN) version 8.6 were used. This specific software stack ensured optimized performance and compatibility between the hardware and the DL libraries, facilitating efficient execution of the training and hyperparameter optimization procedures outlined in Section \ref{subsec:model_selection}.

\end{appendix}

\end{document}